\documentclass[usenatbib]{mn2e}

\usepackage{url}
\usepackage{graphicx}
\usepackage{natbib}
\usepackage{float}
\usepackage{fix2col}
\usepackage{bm}
\footnotesize
\newdimen\minuswidth    %define @ width of minus sign for tables
\setbox0=\hbox{$-$}
\minuswidth=\wd0
\catcode`@=\active
\def@{\kern\minuswidth}
\newdimen\digitwidth    %define ! a one digit width for tables
\setbox0=\hbox{\rm0}
\digitwidth=\wd0
\catcode`!=\active
\def!{\kern\digitwidth}
\normalsize

\title[Interpolation and Prediction in Pulsar Timing]
{Optimal Interpolation and Prediction in Pulsar Timing}
\author[X. P. Deng et al.]
{X. P. Deng,$^{1,3,4}$
W. Coles,$^2$
G. Hobbs,$^3$
M. J. Keith,$^3$
R. N. Manchester,$^3$
\newauthor
R. M. Shannon,$^3$
J. H. Zheng$^1$
\\
$^1$ National Space Science Center, Chinese Academy of Sciences, Beijing~100190, China \\
$^2$ Electrical and Computer Engineering, University of California at San Diego, La Jolla, California, U.S.A. \\
$^3$ CSIRO Astronomy and Space Science, Australia Telescope National Facility, P.O.~Box~76, Epping NSW~1710, Australia \\
$^4$ Graduate University of Chinese Academy of Sciences, Beijing~100190, China
}

\date{}

\begin{document}
\maketitle
\newcommand{\setthebls}{
%                 de-comment this line for double spacing:
%\baselineskip=20pt
}
\setthebls

\begin{abstract}\\

For pulsar projects it is often necessary to predict the pulse phase in advance, for example, when preparing for new observations. Interpolation of the pulse phase between existing measurements is also often required, for example, when folding X-ray or gamma-ray observations according to the radio pulse phase.  Until now these procedures have been done using various ad hoc methods. The purpose of this paper is to show how to interpolate or predict the pulse phase optimally using statistical models of the various noise processes  and the phase measurement uncertainty.

\end{abstract}


\section{Introduction}

Pulsars are rapidly rotating neutron stars. Each rotation of a radio pulsar leads to the characteristic pulse of radiation that can be detected using a radio telescope. Because the rotation of pulsars is exceptionally stable, the time of arrival (ToA) of each pulse can be precisely predicted from a timing model that includes the motions of the pulsar and the Earth and the various propagation effects between the pulsar and the Earth.  Comparing the actual pulse arrival time with the prediction of a timing model, whose parameters are adjusted using a least-squares procedure, is the essence of pulsar timing. The timing model and least squares procedure are implemented in software packages, e.g., \textsc{tempo2}  \citep{G-Hobbs_2006, R-Edwards_2006}.   A preliminary timing model is obtained with the discovery of the pulsar. When a series of observations have been made over a period of years, the timing model can be refined with great precision. The deviations between the actual and the predicted arrival times are known as ``timing residuals.'' Ideally, these residuals would simply reflect the ToA measurement uncertainty, but in fact they will include various un-modeled effects. Residuals often exhibit slow variations which appear to be a red stochastic process, which is referred to as ``timing noise". Timing noise may be caused by many phenomena including those intrinsic to the pulsar  (e.g., Hobbs et al. 2010\nocite{hlk10}, Lyne et al. 2010\nocite{lhk+10} and references therein), effects in the interstellar medium (e.g., You et al. 2008\nocite{yhc+07}), the interplanetary medium (You et al. 2012, in press) or even irregularities in terrestrial time standards (Guinot \& Petit 1991\nocite{gp91}, Petit \& Tavella 1996\nocite{pt96}, Rodin 2008\nocite{rod08} and Hobbs et al. 2010\nocite{hcmc10}).

Being able to remove the timing noise is important for numerous applications. For instance, in order for \cite{G-Hobbs_2004} to measure accurately pulsar proper motions, it was necessary to remove timing noise without affecting any signal with a periodicity close to 1\,yr. This was done using the \textsc{fitwaves} algorithm (described in the Appendix of Hobbs et al. 2004\nocite{G-Hobbs_2004}) which fits a sequence of harmonically related sinusoids to the data, where the period of the highest frequency wave was constrained to be $> 1.5$\,yr. Recent work \citep{W-Coles_2011} showed that improved parameter measurements could be made by obtaining and using a simple, analytic model of the power spectrum of the timing noise in order to calculate the covariance matrices of the red and white noise processes.  With the covariance matrices one can find a linear transformation that whitens the observations. If this transformation is applied to both the observations and the timing model, the least squares problem is greatly simplified. With the covariance matrices the red timing noise can also be interpolated or predicted using a maximum likelihood estimator. It is this interpolation and prediction scheme that we will discuss in the following sections.

In order to obtain the characteristic pulse profile for a pulsar, it is necessary to sum many individual pulses during the observation. For radio observations this is often carried out ``online" by folding the incoming data stream at the known topocentric period of the pulsar. Over short data spans, this can be done with sufficient accuracy by assuming that the timing model is perfect. However, if a pulse profile is obtained from a data set that spans many months or years then it is often necessary to model the timing noise. For instance, obtaining a pulse profile using the Large Area Telescope (LAT) on the {\it Fermi} Gamma-ray Space Telescope requires the gamma-ray photons to be added together according a timing model that must be valid over many years \citep{AA-Abdo_2010_0}. This is commonly done using a timing model from observations of the pulsar obtained with a radio telescope and then ``folding'' the gamma-ray photons into the predicted phase bin according to this model. The radio model gives timing residuals that may be ``whitened'' using the \textsc{fitwaves} procedure \citep{P-Weltevrede_2010}. However, as we demonstrate in subsequent sections, the \textsc{fitwaves} procedure is not theoretically optimal in the sense that a maximum likelihood estimator is optimal, nor can it be extrapolated past the end of the radio data (or back-extrapolated before the start of the radio data). \textsc{Fitwaves} also requires an arbitrary choice of the number of harmonics to include in the fitting procedure. Too few harmonics implies that not all the features in the timing residuals are completely modelled. Too many harmonics leads to an unphysical model for the timing residuals (particularly when large gaps exist in the data). %In this paper we describe a means that can optimally interpolate and extrapolate the timing residuals without the need for \textsc{fitwaves}. 

Prediction of the pulse phase into the future is required for real-time folding of pulsar data (a very widely used technique).  This is typically carried out by extrapolating the pulsar timing model into the future and ignoring the expected variations due to the pulsar timing noise (the extrapolated timing model is typically described using a Chebyshev polynomial expansion for efficient implementation in digital hardware; see \citealt{G-Hobbs_2006} for details). 

Another example of a practical use of the requirement to predict pulse phase into the future is in order to help navigate space probes in the Solar System \citep{S-I-Sheikh_2005, S-I-Sheikh_2006, bernhardt2010,bernhardt2011}. Comparison of pulse arrival times at the satellite and the expected pulse arrival times given a pulsar timing model can be used to determine the position of the satellite. However, it is expected that any such system will only have the sensitivity to observe pulsars with high X-ray flux density; such pulsars are typically young and have residuals that exhibit significant timing noise (e.g., Gavriil \& Kaspi 2002)\nocite{F-Gavriil_2002}. Optimal extrapolation of the timing residuals is therefore essential for such navigation.

In this paper, we describe how to use the covariance matrices of the red and white noise processes to both interpolate and extrapolate pulsar timing residuals, without the need for \textsc{fitwaves}.  In Section 2, we outline the theory.  In Section 3, we carry out simulations to demonstrate the effectiveness of our technique. In section 4, we compare the new technique with \textsc{fitwaves} in timing noise modelling of radio data from observations of the Vela pulsar, B0833$-$45 for folding of {\it Fermi} gamma ray observations and outline the advantage of the new algorithm in space-craft navigation.  In Section 5, we apply our algorithm to a simulated data set that includes a glitch event, use simulations to establish the sensitivity of the method to errors in parameterising the noise and use real data from observations of PSR~J1939$+$2134 to compare the new algorithm with and without quadratic removal. We conclude the paper in Section 6.

\section{The Maximum Likelihood Filter}

The theory of optimal interpolation and extrapolation of stationary time series was first discussed by Wiener \citeyearpar{N-Wiener_1964} and the corresponding algorithm has become known as a Wiener filter.  However normal Wiener filters are not applicable to pulsar timing residuals because the measurement uncertainty is not stationary and the sampling times are irregular.  However, Wiener filters are maximum likelihood estimators (MLE) and we can obtain an MLE for pulsar timing residuals directly as shown below if we know the covariance of the red and white noise processes.  This is effectively a generalised Wiener filter.

Here we wish to estimate a red noise process ($\bm{s}$), from irregularly spaced observations ($\bm{o}$) of the red noise plus measurement error ($\bm{n}$), i.e. $\bm{o = s + n}$, when both processes are gaussian.  The likelihood function,  i.e. the logarithm of the probability density, for N observations (neglecting the normalizing constants) is given by

\begin{equation}
\bm{s}^\mathrm{T}\mathbf{C}_\mathrm{s}^{-1}\bm{s}+\bm{n}^\mathrm{T}\mathbf{C}_\mathrm{n}^{-1}\bm{n}=\bm{s}^\mathrm{T}\mathbf{C}_\mathrm{s}^{-1}\bm{s}+\bm{(o-s)}^\mathrm{T}\mathbf{C}_\mathrm{n}^{-1}\bm{(o-s)}
\end{equation}
Here $\mathbf{C}_\mathrm{n}$ is the covariance matrix of the measurement error and $\mathbf{C}_\mathrm{s}$ is the covariance matrix of the red noise. To find the MLE estimator for $\bm{s}$ given observations $\bm{o}$, we set the gradient of the likelihood function with respect to $\bm{s}$ to zero and solve the resulting linear system,
\begin{equation}
2\mathbf{C}_\mathrm{s}^{-1}\bm{s}-2\mathbf{C}_\mathrm{n}^{-1}\bm{(o-s)}=0\\
\end{equation}
\begin{equation}
(\mathbf{C}_\mathrm{s}^{-1} + \mathbf{C}_\mathrm{n}^{-1})\bm{s}=\mathbf{C}_\mathrm{n}^{-1}\bm{o}.\\
\end{equation}

$\mathbf{C}_\mathrm{n}$ is a diagonal matrix so it is trivial to invert, but the inversion of $\mathbf{C}_\mathrm{s}$ can be unstable so the Cholesky decomposition should be used if it is necessary to invert it. Equation 3 can be solved directly or it can be simplified first by pre-multiplying both sides of $\mathbf{C}_\mathrm{n}$ or $\mathbf{C}_\mathrm{s}$. We pre-multiply both sides by $\mathbf{C}_\mathrm{s}$ to avoid the need for inverting $\mathbf{C}_\mathrm{s}$. 

This solution provides an estimate of the timing noise $\bm{s}$ at every observed sample. However the MLE is not as restrictive. We can solve for the signal that maximizes the likelihood function, even at locations where we do not have a sample $\bm{o}$. In this case the likelihood function will include $\bm{s}$ terms that are not matched by $\bm{o}$ terms. Thus in equation 2 the term $\mathbf{C}_\mathrm{n}^{-1}\bm{(o-s)}$ will include only the subset of $\bm{s}$ which matches $\bm{o}$. We simplified the system by augmenting the $\bm{o}$ vector with terms of value zero terms matching the unmatched $\bm{s}$ terms. We also augmented $\mathbf{C}_\mathrm{n}^{-1}$ with terms of value zero along the diagonal, corresponding to the augmented terms of $\bm{o}$. With these extensions to $\bm{o}$ and $\mathbf{C}_\mathrm{n}^{-1}$ the system can be reduced to the form in Equation 3 and solved as discussed earlier.

\section{Testing the algorithm}

\begin{figure}
\includegraphics[height=8cm,angle=-90]{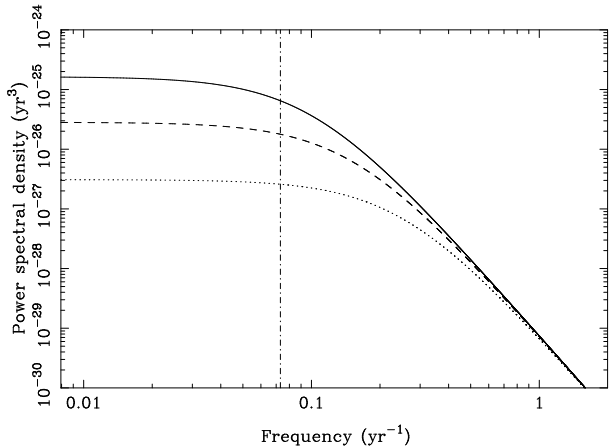}\\
\caption{Power spectral densities for the three types of ``timing noise''. The solid line is defined by $A=7.6\times10^{-30}, \alpha=4.3333$, $f_\mathrm{c}=0.1\mathrm{yr}^{-1}$ and $f_\mathrm{o}=1\mathrm{yr}^{-1}$. The dashed line with $A=7.6\times10^{-30}, \alpha=4.3333$, $f_\mathrm{c}=0.15\mathrm{yr}^{-1}$ and $f_\mathrm{o}=1\mathrm{\mathrm{yr}}^{-1}$. The dotted line with $A=7.6\times10^{-30}, \alpha=4.3333$, $f_\mathrm{c}=0.25\mathrm{yr}^{-1}$ and $f_\mathrm{o}=1\mathrm{yr}^{-1}$. The vertical dot-dashed line is at a frequency of $1/T_{\rm span}$.}\label{fg:specDensity}
\end{figure}

\begin{figure}
\includegraphics[height=8cm,angle=-90]{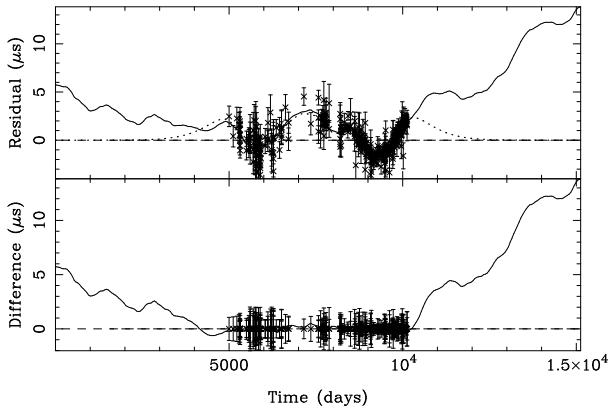}
\caption{The upper panel shows one realisation of simulated timing residuals (with same error bars as real data). The irregular solid line shows the extrapolation of the data for 5000\,d before and after the end of the actual sampling. The dotted line indicates the prediction of the timing noise using our algorithm. The lower panel shows the difference between the simulation and the prediction with the error bars overlaid where sampled.}\label{fg:result1}
\end{figure}

\begin{figure}
\includegraphics[height=8cm,angle=-90]{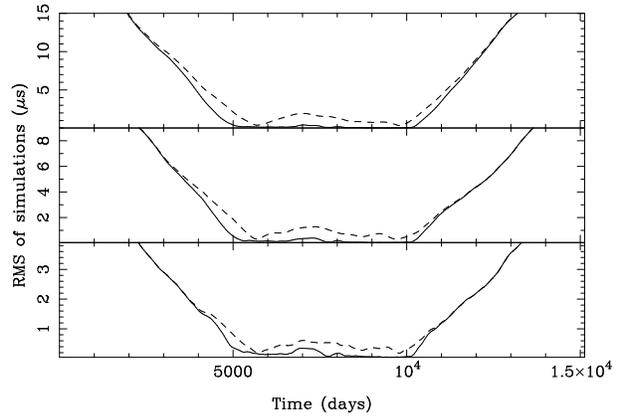}
\caption{50 realisations of each type of timing noise shown in Figure 1. The top panel contains the results for the strongest (defined as the highest amplitude noise in Figure 1) red noise, the middle panel contains the intermediate-level noise and the lowest panel contains the weakest red noise. The rms of these realizations for each sample is shown as the dashed lines.  The rms difference between the known simulated noise and the interpolated/extrapolated predictions are shown as the solid line.}\label{fg:rms}
\end{figure}

In order to test our algorithm we first simulated data sets that contain both white and red noise. In order to obtain realistic sampling and ToA uncertainties we used observations of PSR~J1713$+$0747 obtained using the Parkes radio telescope from MJD 49421.9 to MJD 54546.8 \citep{J-P-W-Verbiest_2009} as an example of a ``typical millisecond pulsar'' data set. With this sampling, we simulated a data set of timing residuals with red noise and added white noise to these generated observations. We carried out simulations with three different forms of timing noise. In all cases the white noise level remained the same. The simulated spectrum model has the form $P(f)=A/[(f_\mathrm{c}/f_\mathrm{o})^2+(f/f_\mathrm{o})^2]^{(\alpha/2)}$.  $A$ is the amplitude of the red timing noise, $f_c$ is a corner frequency that ensures that the red noise model turns over at low frequencies, $f_\mathrm{o} = 1$yr$^{-1}$ and $\alpha$ is the spectral exponent of the red noise. The spectral densities of the three timing noise types are shown in Figure~\ref{fg:specDensity}.   We define these three noise types as having ``strongest'', ``intermediate-level'' and ``weakest'' red noise depending on the amplitude of the red noise process. Note we choose the spectral exponent of $\alpha = 4.3333$ to represent the red noise as this is expected if the timing residuals were dominated by the effect of a stochastic background of gravitational waves (e.g., Sesana, Vecchio \& Colacino 2008)\nocite{svc08}\footnote{Note that the characteristic strain of a gravitational wave background has the form $h_c(f) = A_gf^{\alpha_g}$ where $\alpha_g = -2/3$.  The power spectral density of the induced timing residuals has the form $P = Af^{2\alpha_g-3}$.}. Similarly steep spectral exponents are also typical of other forms of timing noise such as residuals caused by unexplained rotational irregularities, glitch recovery or unmodelled interstellar medium effects \citep{cordes2010measurement}. The cutoff frequency was chosen to lie close to 1/$T_{\rm span}$ (where $T_{\rm span}$ is the data span of the actual observations), shown as the vertical dot-dashed line. 

For each of the three spectral models we computed a single daily-sampled realization of the timing noise. In order to test our extrapolation procedure, each simulation had a data span of the actual observations, but with an extra 5,000 days added to the start and to the end of the data span. We require multiple realizations of this red noise. To do this we created one data set that was 50 times longer than the required data length. We created one long data set so that we could simulate the low frequencies ($f < 1/T_{\rm tot}$ where $T_{\rm tot}$ is the required data span) with a simple discrete Fourier transform. This data set can then be subdivided into blocks of length $T_{\rm tot}$ to give different realisations of the random process.  For the central region that corresponds to the actual observations, we interpolated the timing noise to the actual sample times using a constrained cubic interpolator. We subsequently added white noise to these interpolated samples at the level given by the ToA uncertainties for each observation. We also regularly sample the data every $T_{\rm span}/100 \sim 50\,d$ over the entire data span, but do not add white noise to these data as the regularly spaced samples are used to provide a comparison between the actual timing noise and the interpolated timing noise. In order to simulate the effect of fitting for the pulsar timing model, we fitted, and subsequently removed, a quadratic polynomial to the data that represented the actual observations.

We use our algorithm to determine the signal using the same gridding as the regularly sampled data described above. In this way we could compare the results of our interpolation, extrapolation and back-extrapolation with the known red noise that was simulated. For these simulations, we did not obtain the covariance functions from the data; instead we used the known analytical power spectra to compute the covariance functions and the corresponding matrices. The results for the intermediate timing noise signal (the middle spectrum in Figure~\ref{fg:specDensity}) are shown in Figure~\ref{fg:result1}. The top panel shows the simulated residuals for one realization of the red noise (shown with error bars). The timing noise realization is shown as the irregular solid line. Note that the timing noise diverges from the timing model after the range of the quadratic fit. Our algorithm leads to the interpolation and extrapolation that is shown as a dotted line. The difference between the simulated red noise and the predicted function is shown in the lower panel of Figure~\ref{fg:result1}. Error bars representing the white noise level are overlaid, at the locations of the actual sampled data points. During the range of the observations, the interpolation works well. Discrepancies between the interpolation function and the simulated red noise level are smaller than the white noise level. Outside of the observing span, the extrapolated function converges to the prediction of the timing model (i.e., predicted residuals of zero). The discrepancies between the extrapolated function and the simulated data grow as the extrapolation time increases. The rms of the discrepancies between the model and the known signal measured over all 50 realizations of all three types of timing noise is shown as the solid line in Figure~\ref{fg:rms}. In all cases the rms of the discrepancies converge to the rms of the timing noise (shown as the dashed line in Figure~\ref{fg:rms}).

\section{Applications}

\subsection{Timing noise modelling and removal}

\begin{figure*}
\includegraphics[width=7cm,angle=-90]{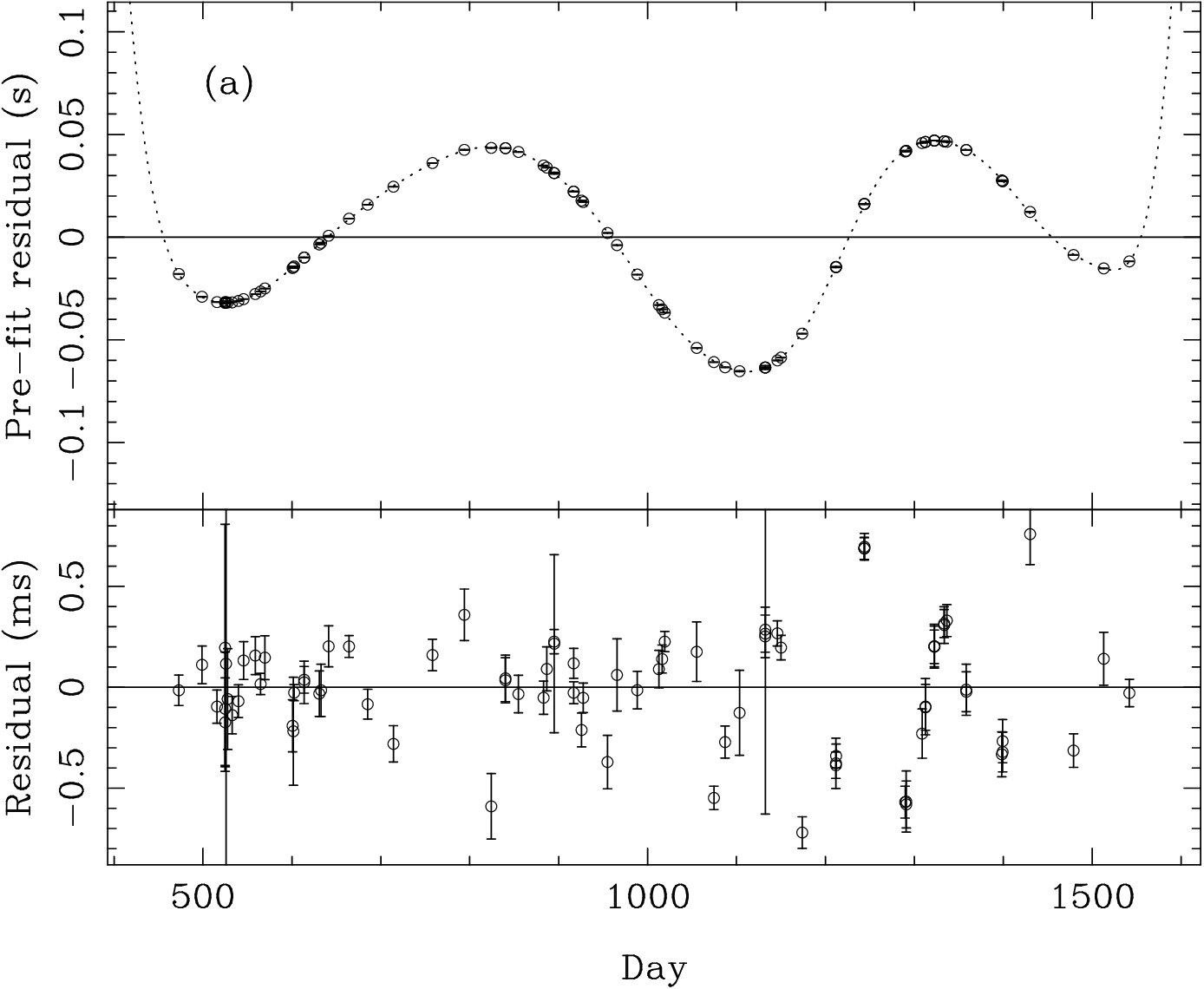}
\includegraphics[width=7cm,angle=-90]{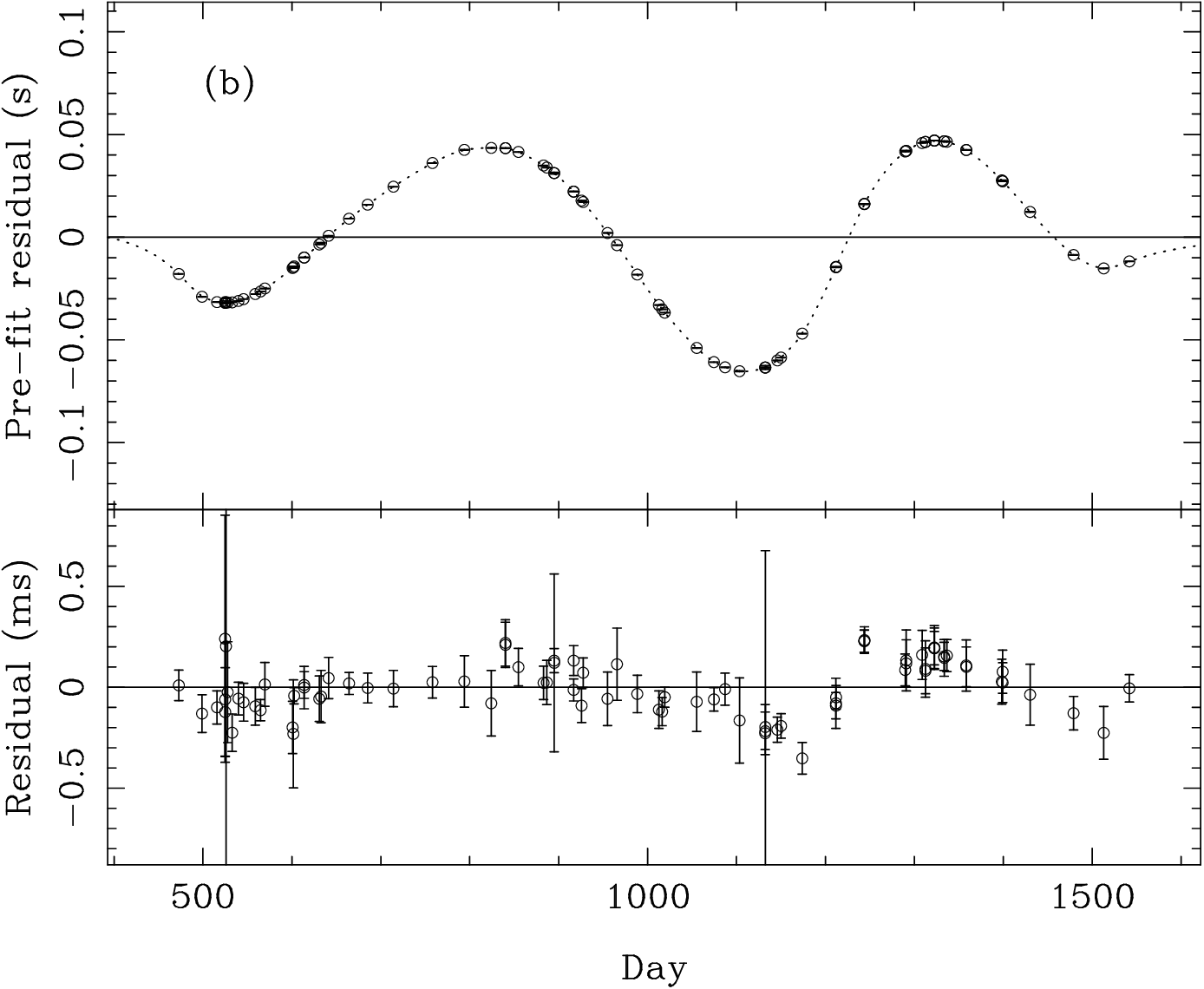}
\includegraphics[width=7cm,angle=-90]{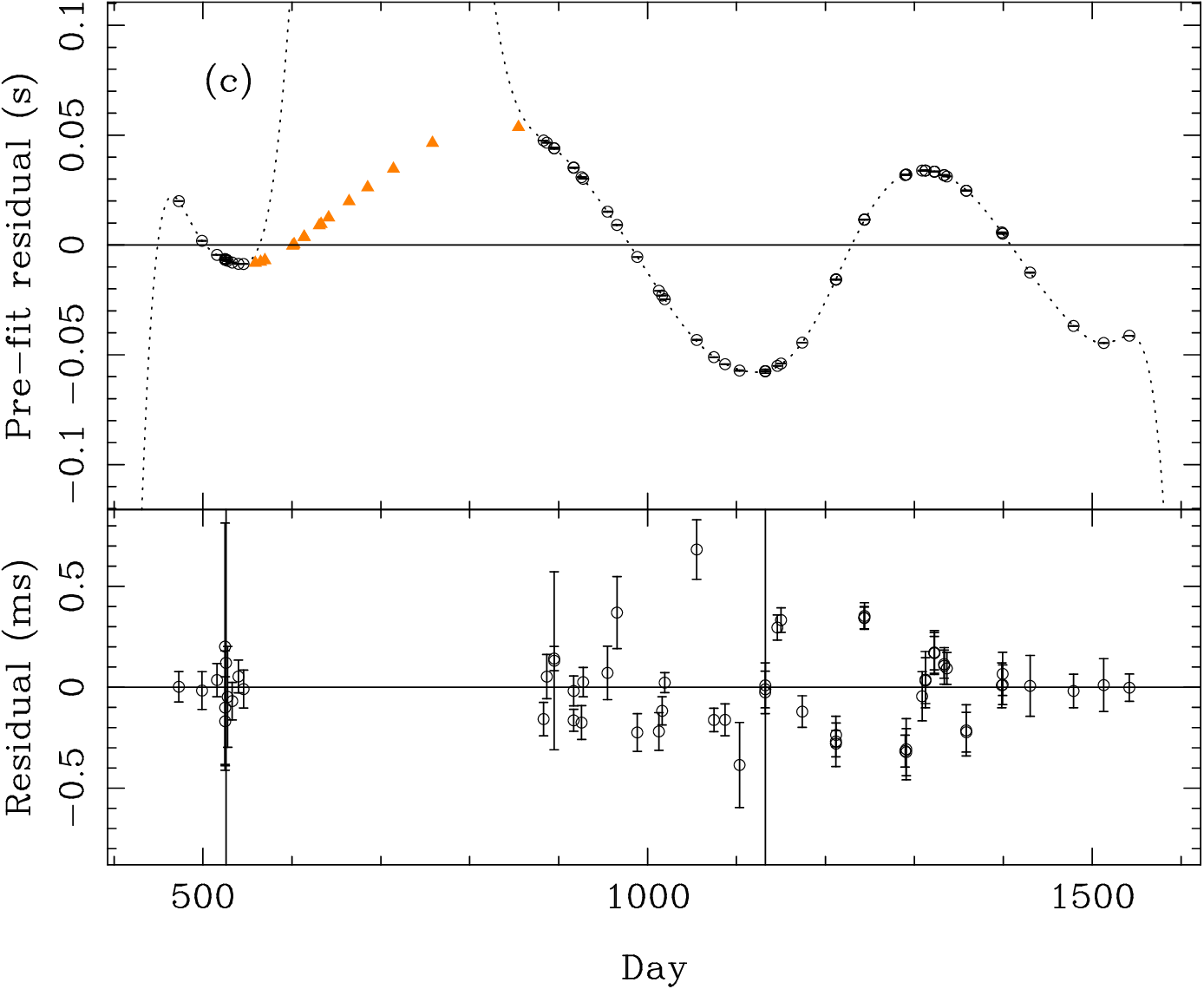}
\includegraphics[width=7cm,angle=-90]{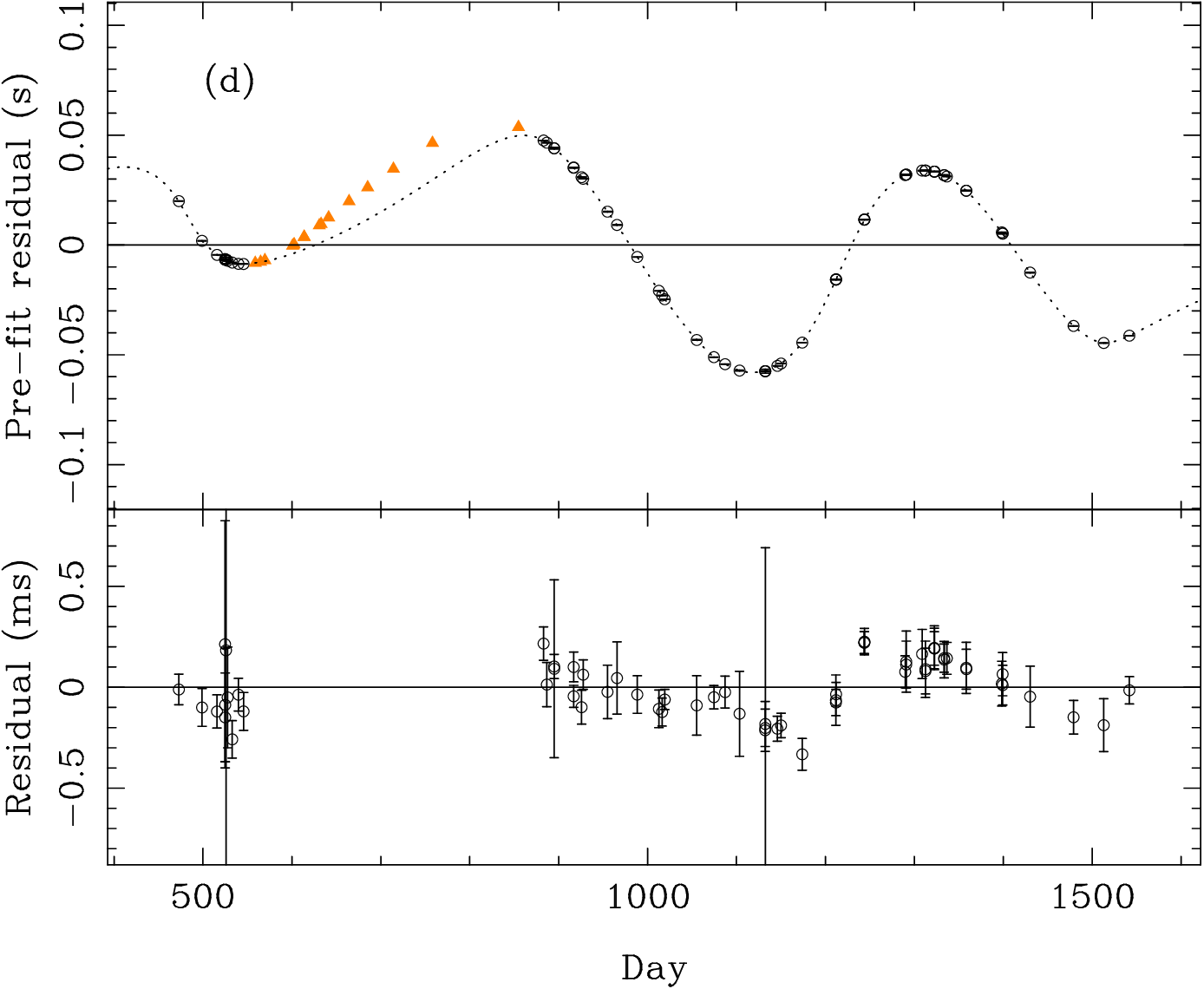}
\includegraphics[width=7cm,angle=-90]{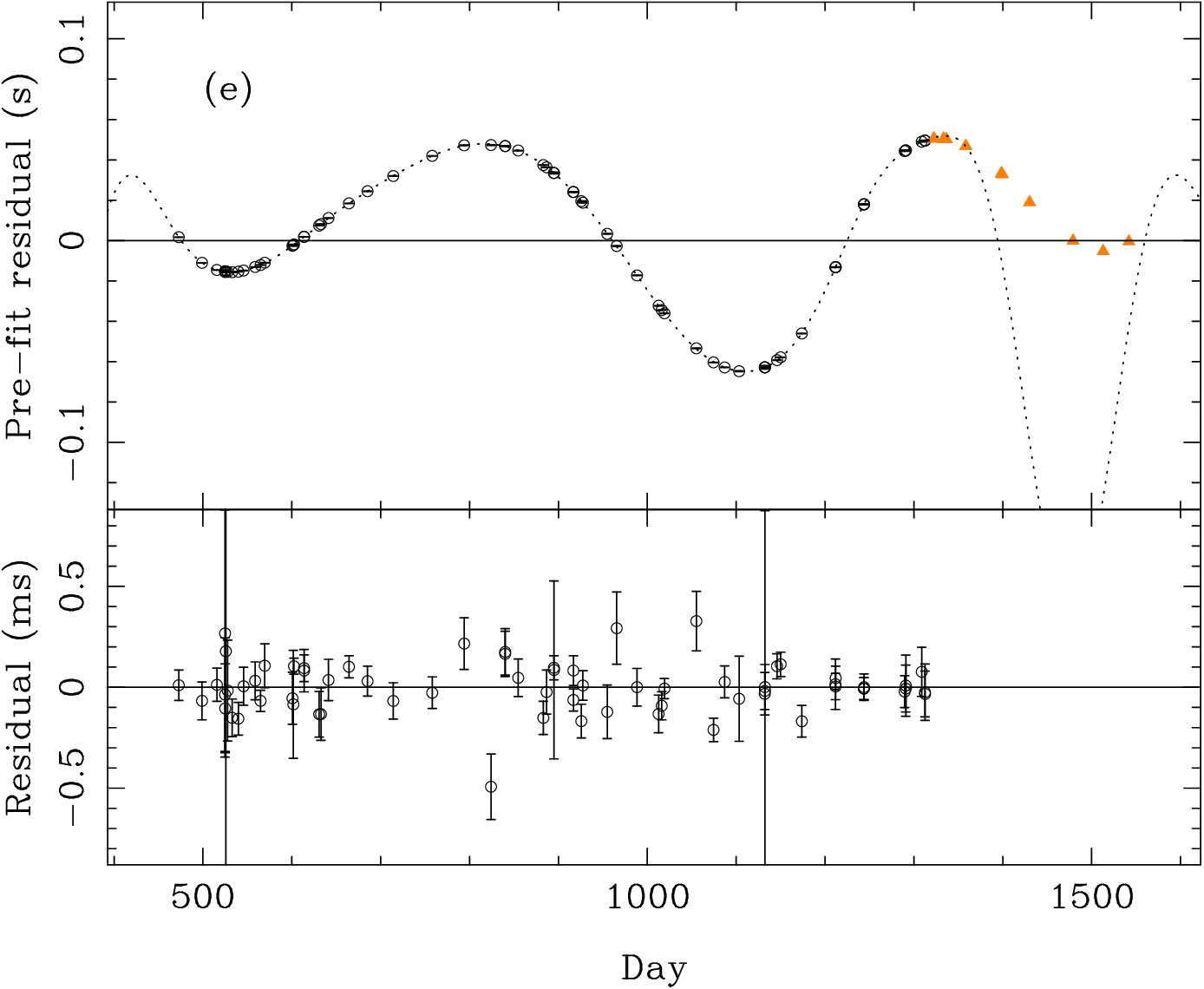}
\includegraphics[width=7cm,angle=-90]{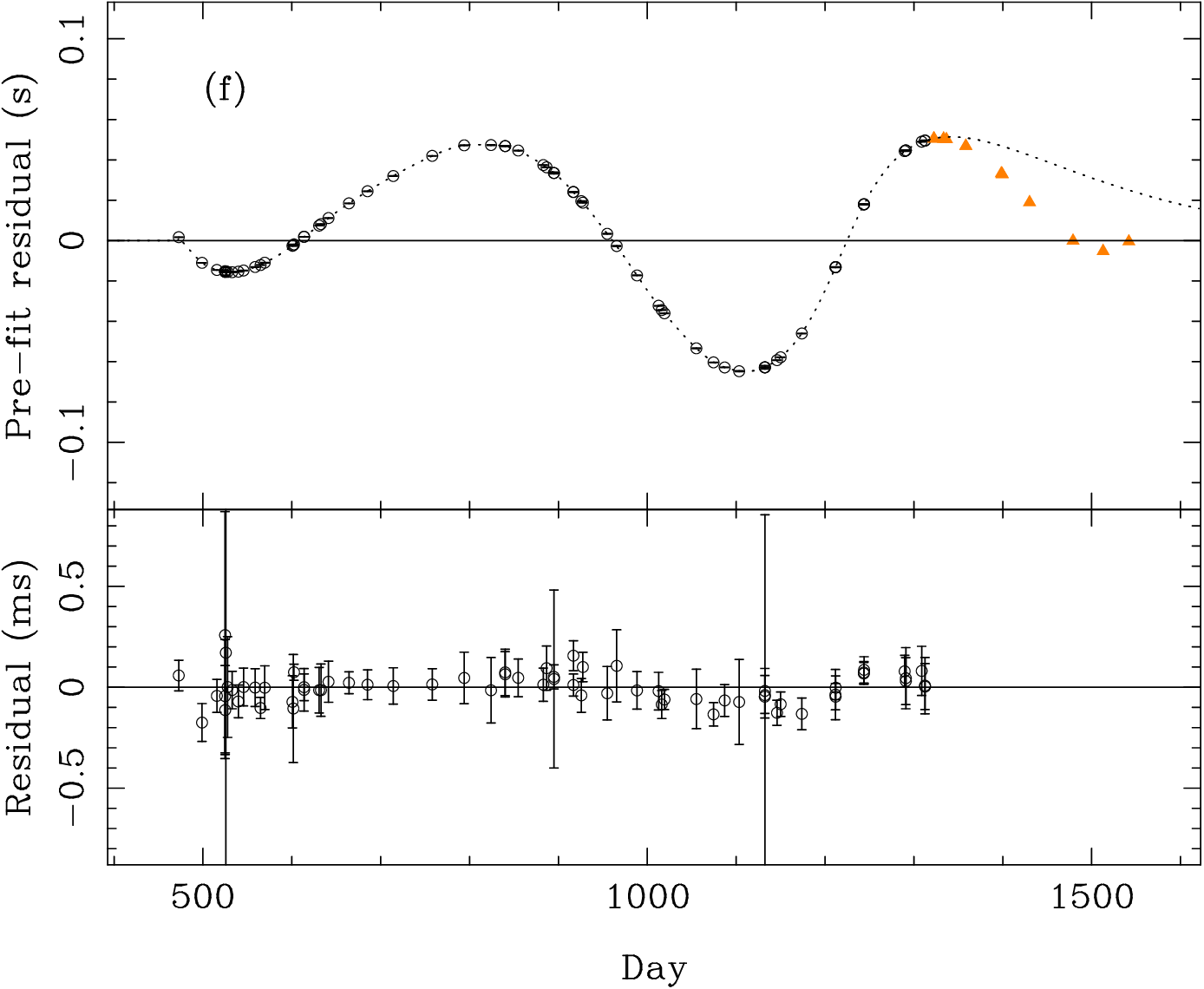}
\caption{Models of the timing residuals for the Vela pulsar. The left-hand panels (a,c,e) show models of the residuals using the \textsc{fitwaves} procedure. The top part of each panel contains the original timing residuals and the model (dotted line). The lower part of each panel shows  the difference between the actual residuals and the model. The right-hand panels (b,d,f) show the same, but using the interpolation/extrapolation procedure described in this paper. The panels (c,d) have data removed before the fitting, interpolation and extrapolation. These ``removed'' observations are shown with the small triangle symbols. The panels (e,f) have the most recent 220\,d of data have  removed from the analysis.}\label{fg:vela}
\end{figure*}

In order to test our algorithm using a real pulsar data set that exhibited a significant amount of timing noise we made use of radio observations of the Vela pulsar (PSR B0833$-$45) from the Parkes 64-m radio telescope \citep{P-Weltevrede_2010}. These data were taken between MJD 54689 and 55758 mainly to support the {\it Fermi} space-craft mission. In order to determine the covariance function of the red noise, we used the Cholesky algorithm described in \cite{W-Coles_2011} and implemented as the \textsc{spectralModel} plugin to \textsc{Tempo2}. We used this resulting covariance function to obtain precise astrometric and pulse parameters for the Vela pulsar and formed the post-fit timing residuals. These are shown in the top panels of Figure~\ref{fg:vela}a and \ref{fg:vela}b (the ToA uncertainties are smaller than the size of the symbol). In common with most analyses of these data, we used the \textsc{fitwaves} algorithm (with 10 harmonics) to model the timing noise. This model is shown as the dotted line in the top panel of Figure~\ref{fg:vela}a and these post-fit residuals, once this model is removed, are shown in the bottom panel of Figure~\ref{fg:vela}a. Note that the data are modelled well during the observations, but the \textsc{fitwaves} function diverges past the end of the timing residuals and therefore cannot be used to extrapolate the behaviour. 

For comparison, we have applied our procedure and provided a model for the timing noise with a grid spacing of 13\,d. The model is subsequently interpolated, using a linear interpolation, to determine the pulse phase at the exact time of each observation. Our model is shown as the dotted line in the top panel of Figure~\ref{fg:vela}b. We note that this method models the data well during the observations and converges to the predictions of the timing model about one hundred days after the end of the data span. The post-fit residuals clearly have a lower rms timing residual than the post-fit residuals obtained with the \textsc{fitwaves} algorithm.

A significant problem with the \textsc{fitwaves} algorithm is shown in the top panel of Figure~\ref{fg:vela}c. For this Figure we have removed a few hundred days of radio observations near the start of the data span (these ``removed'' data points are indicated in the Figure using small triangle symbols, but are not included in the analysis). The \textsc{fitwaves} model (shown as the dotted line) interpolates poorly during the data gap. This can lead to significant problems with folding, e.g., gamma-ray photons that were observed during this interval. In contrast, our algorithm (Figure~\ref{fg:vela}d) provides an interpolation that more closely represents the original observations during the data gap. 

\subsection{Spacecraft navigation}

It may be possible to use observations of pulsars to assist with navigation of space probes throughout the Solar System. This method relies on the ability to predict the arrival times of pulses precisely and accurately. However, previous work \citep{S-I-Sheikh_2005, S-I-Sheikh_2006,bernhardt2010,bernhardt2011} has assumed that the pulse arrival times can be modelled using a standard pulsar timing model. Even though such a timing model can be ``whitened'' using existing techniques (i.e., the \textsc{fitwaves} procedure), it is likely that, for any navigational purpose, it will be necessary to extrapolate the timing model. As shown above, the \textsc{fitwaves} procedure cannot be used for such extrapolation. In Figures~\ref{fg:vela}e and \ref{fg:vela}f we have shortened the actual radio observations of the Vela pulsar by 220 days and re-fitted for the pulsar's pulse frequency and its first time derivative.  We then use \textsc{fitwaves}  (Figure~\ref{fg:vela}e) and our new algorithm (Figure~\ref{fg:vela}f) to predict the timing residuals and compare with the actual data.  The actual data included in the analysis are shown as circles.  The data that have been removed are indicated using triangle symbols.  Clearly the extrapolation is not perfect with our method.  After only a few tens of days the prediction has diverged from the actual data by an amount larger than the white noise error bars.  However, in contrast to the \textsc{fitwaves} procedure, the prediction is never worse than the rms of the actual timing residuals.

The error in determining the space-probe position is related to the error in predicting the pulse arrival times. However, without prior information, positional determination can only be obtained by comparing the actual and predicted arrival times for three or more pulsars. If the predictions for one, or more, of these pulsars is incorrect (for instance, due to poor extrapolation) then it may be impossible to obtain a solution for the position of the space-probe. 

\section{Discussion}

\subsection{Glitch events}

%The algorithm presented here is only optimal if the timing noise is stationary.
The algorithm presented here is correct if the covariance matrix exists. It does not have to be stationary (or wide-sense stationary). However the scheme for estimating the covariance matrix presented by \citet{W-Coles_2011} assumes that the red noise is stationary. If the noise is non-stationary, for example if small glitch events exist in the residuals, then this information should be included in the timing model to the extent possible.  However, in some cases, the glitch event may not be easily identifiable.  Here we consider whether the covariance matrix can deal with weakly non-stationary observations. 

\begin{figure}
\includegraphics[height=8cm,angle=-90]{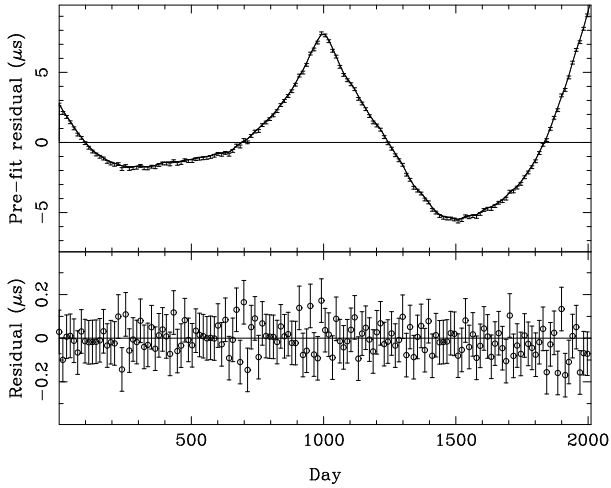}
\caption{Interpolation of timing residuals that include a glitch event. The upper panel shows the timing residuals with the interpolated function shown as a solid line.  The lower panel shows the difference between the actual data and the interpolation.  The error bars are defined by the 100\,ns of white noise simulated for this data set.}  \label{fg:glitchInterp}
\end{figure}

We have simulated a data set with regularly spaced observations at 14\,d intervals and white noise with an rms timing residuals of 100\,ns.  The exact amount of white noise is not important; this value was chosen so that the effect of the glitch is large compared with the white noise level.   We have added a moderate red noise signal to simulate the effect of timing noise and included a small glitch (corresponding to a frequency change of $1\times10^{-12}$\,Hz) exactly at the centre of the data span.  We have formed timing residuals, obtained a spectral model of the noise and used our algorithm to interpolate the data points. The result is shown in Figure~\ref{fg:glitchInterp}.  The difference between the interpolated timing residuals and the simulated data is less than the white noise level, demonstrating that the interpolation procedure correctly models the timing residuals.

\subsection{Sensitivity to errors in the parameterisation of the noise}
\begin{figure}
\includegraphics[height=8cm,angle=-90]{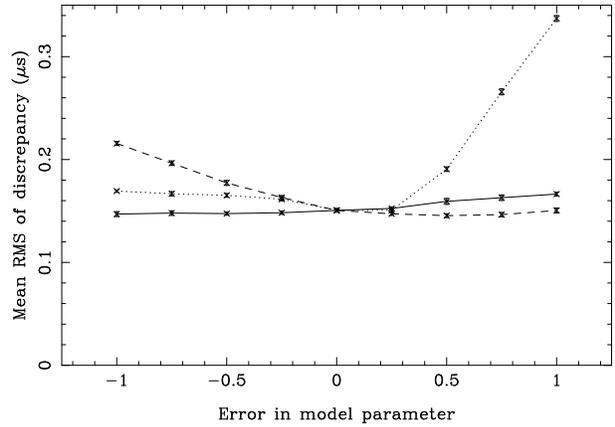}
\caption{Sensitivity of the measurements to changes in the noise model.   The dotted, dashed and solid lines show the sensitivity to changes in the parameterisation of the spectral component, $\alpha$, corner frequency, $f_\mathrm{c}$ and the white noise, $\sigma_\mathrm{w}^2$, respectively.  These are represented on the horizontal axis by  $\alpha-\alpha_0$, $\log_{10}(f_c/f_{c0})$ and $\log_{10}(\sigma_w^2/\sigma_{w0}^2)$  where $\alpha_0$, $f_{c0}$ and $\sigma_{w0}^2$ are the known values for these parameters.} \label{fg:noiseError}
\end{figure}

Our algorithm allows pulsar timing residuals that exhibit red noise to be modelled by interpolation or extrapolation. In contrast to previous methods, such as using the \textsc{fitwaves} algorithm or fitting high-order polynomial expansions, the extrapolated model converges on the prediction of the pulsar timing model.

Our MLE method is optimal only if the statistical properties of the noise is known, so it is important to establish its sensitivity to errors in the parameterisation. We have repeated the simulations described in Section 3, but here we used incorrect estimates of the spectral model and the white noise covariance. The timing noise simulated corresponds to the lowest spectrum in Figure 1. As before, we kept the ToA uncertainties and the sampling the same as for the actual observations of PSR~J1713$+$0747 (note, here our data sets were not enlarged by 10,000 days as before). Our procedure is as follows: 
\begin{itemize}
\item{Make 50 realizations of the noise (as in Section 3).}
\item{Measure the rms of the interpolation error between the different realisations for each sample. These are then averaged to give a single value.}
\item{This entire procedure is repeated 10 times. The mean and the rms of the mean are recorded.}
\item{The parameters of the model are changed (modifications are made to $\alpha$, $f_c$ or $\sigma_w^2$ in turn keeping the other two parameters fixed at their correct value) and the entire procedure repeated. }
\end{itemize}

The mean rms values are shown as a function of the error in the three model parameters as the three lines in Figure~\ref{fg:noiseError}. The error bars (which are generally smaller than the symbol size) represent the rms of the mean. 

The largest discrepancies are seen with an incorrect determination of the corner frequency $f_c$. The algorithm is very insensitive to errors in the spectral exponent, $\alpha$. In all cases the discrepancies are significantly smaller than the mean rms of the white noise (0.69$\mu$s) and the mean rms of the red noise (0.42$\mu$s).  
We note that the parameter value that leads to the lowest mean rms in Figure~\ref{fg:noiseError} does not correspond exactly to the known model parameter.  This is probably because the red noise process is not stationary after the quadratic polynomial removal.  We have also studied the impact of incorrect model parameters on extrapolating the timing residuals.     For incorrect model parameters the duration for which the timing residuals can be extrapolated will be reduced when compared with the correct model parameters.  The maximum extrapolation duration will depend upon the size of the error in the parameters, the white noise level and the requirements of the specific application.  However, even with significant errors in the model parameters, the variation in the discrepancy between the predicted residuals and the known simulated red noise will be smaller than variations in different realisations of the red noise. The technique is therefore insensitive to errors in the model parameters and this will not affect our method for actual data sets. 

\subsection{Quadratic removal of timing noise}
%\begin{figure}
%\includegraphics[width=6cm,angle=-90]{fig7.ps}
%\caption{The effect of quadratic removal on the timing noise rms. The top panel show the rms of original red noise, the second panel shows the rms of red noise with regularly sampled quadratic removal, the third panel shows the rms of red noise with irregularly sampled quadratic removal.}
%\end{figure}

%During pulsar timing analysis, a quadratic function is always fitted and removed from the pulsar timing residuals. This leads to numerous effects. Firstly, the resulting rms of each sample with multiple realisations (as in Section 3) may vary across the data span. To demonstrate this effect we used the lowest spectrum model in Figure 1 to simulate multiple red-noise realisations as in Section 3. In the top panel of Figure 7 we show the rms of the samples without quadratic removal (the data here are the original daily spaced red noise as in Section 3). In this case the values are not very dependent on the specific sample. In the middle panel we show the same, but where the data have been sampled uniformly and a quadratic function has been fitted and removed (the sampling interval here is about 51.2 days). In this case, the rms values are lower, but still independent of the actual sample. In the lowest panel, we sample the data in the same manner as the actual PSR~J1713$+$0747 observations and fit and remove a quadratic polynomial from the unevenly sampled data set. In this case the rms values vary across the data span. 

\begin{figure*}
\includegraphics[height=8cm,angle=-90]{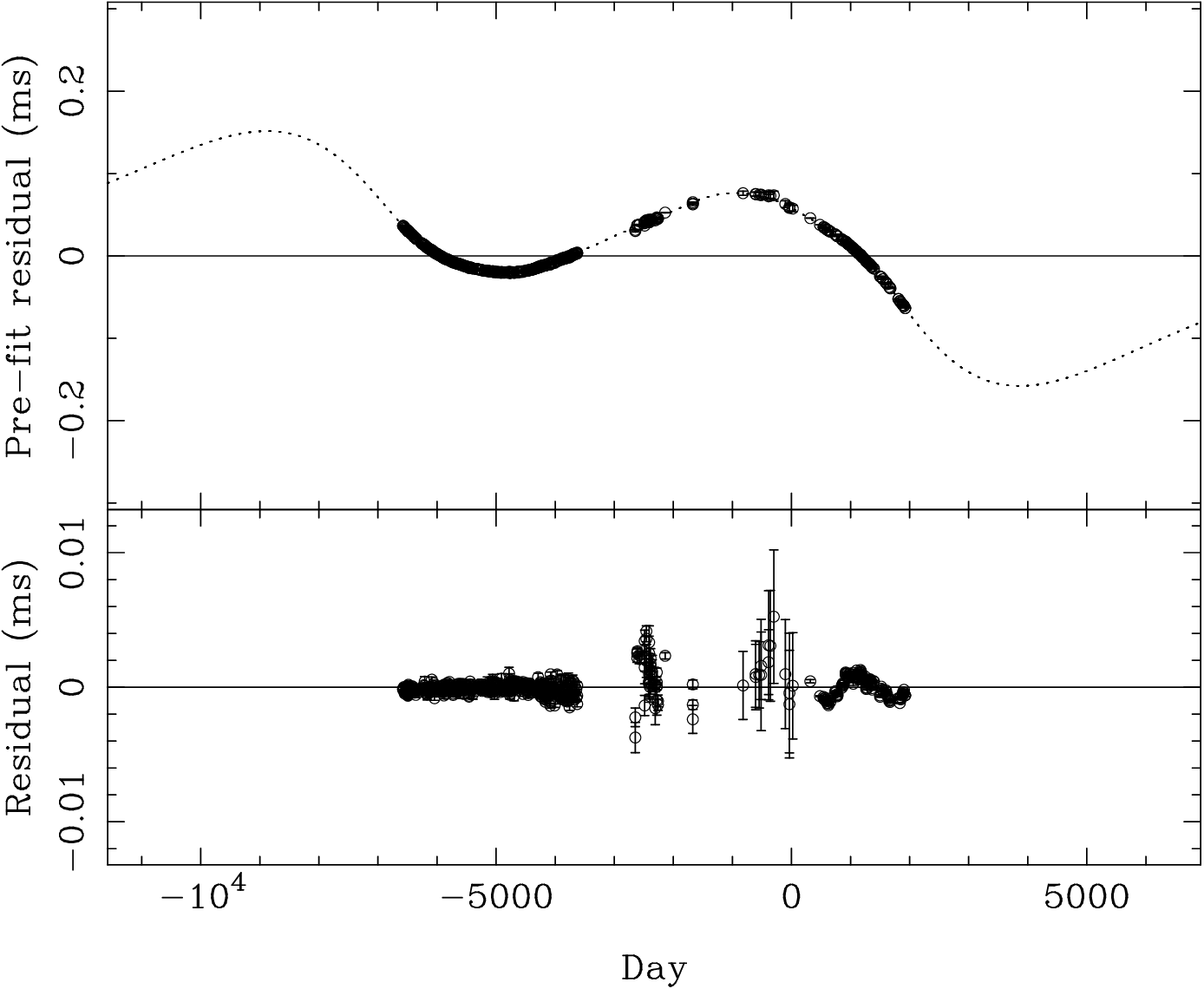}
\includegraphics[height=8cm,angle=-90]{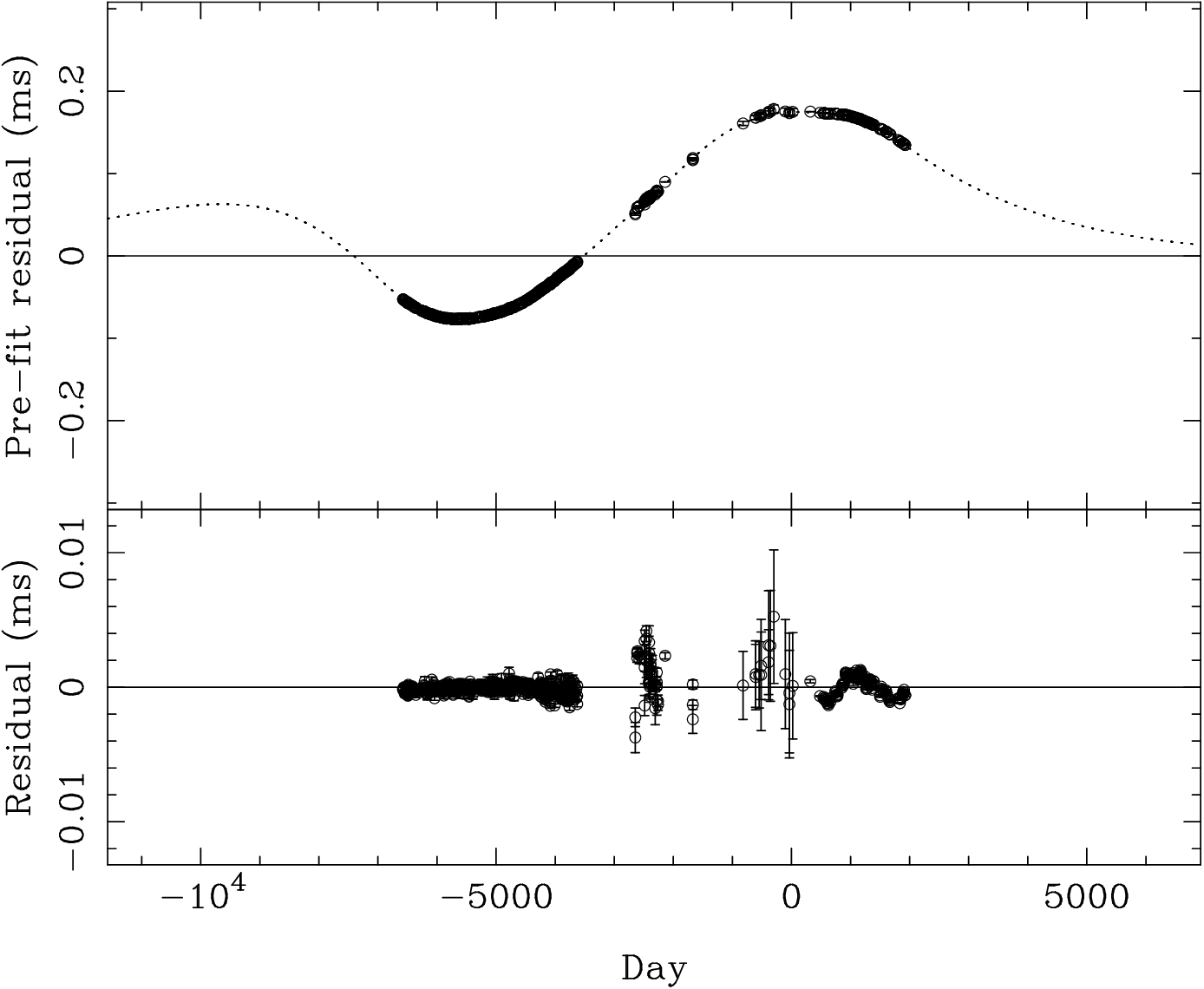}
\caption{Interpolation and extrapolation of the PSR~J1939+2134 timing residuals (dotted line). The left-hand top panel shows the residuals after fitting for the pulse frequency and its first time derivative without using the \citet{W-Coles_2011} algorithm. The right-hand top panel shows the same after using the \citet{W-Coles_2011} algorithm. The lower parts of each panel show the difference between the residuals and the interpolated function during the data span.}\label{fg:1939}
\end{figure*}

The observed pulsar timing residuals depend on the method used to fit for the pulsar's pulse frequency and its first derivative.   In Figure~\ref{fg:1939} (which is based on Figure 1 of \citealt{W-Coles_2011}) the upper section of the left-hand panel shows the timing residuals for PSR~J1939$+$2134 obtained using the Parkes radio telescope. The residuals are significantly red and, after fitting and removing a quadratic polynomial,  take the form of a cubic polynomial. However, if this fit is carried out using the generalised least-squares-fitting procedure \citep{W-Coles_2011} then the resulting residuals still exhibit a linear and quadratic term (this is shown in the right-panel of Figure~\ref{fg:1939}). We have used our algorithm to interpolate and extrapolate these residuals. The results are shown as the dotted line in the two panels of Figure~\ref{fg:1939}. The difference between the model and the residuals are shown in the lower sections of each panel. In both cases the interpolation works well and converges on the predictions of the timing model. However, because of the different post-fit parameters, the timing models are different in the two cases. 

\begin{figure}
\includegraphics[height=8cm,angle=-90]{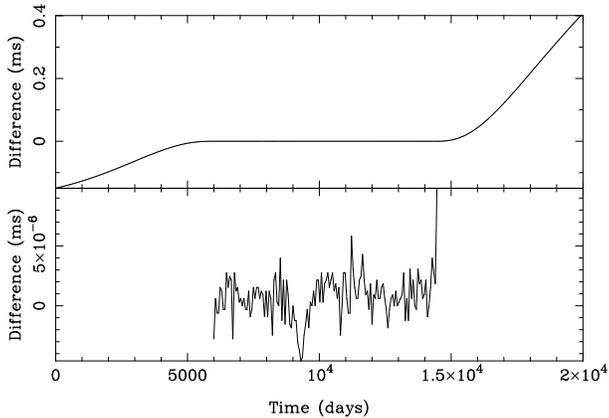}
\caption{Comparison of two models for PSR~J1939$+$2134. The top panels shows the difference between determinations of absolute arrival times given two models for PSR~J1939$+$2134, shown in Figure 7. The bottom panel is a zoom-in of the upper panel during the region where actual data exist.}\label{fg:absTime}
\end{figure}

If this pulsar is being used for navigational purposes then measurements of absolute pulse arrival times are necessary.  From the two panels of Figure~\ref{fg:1939}, it may be expected that the two solutions and extrapolations will lead to discrepant measurements of the absolute pulse arrival times.  In order to test this, we have determined pulse arrival times using both models shown in Figure~\ref{fg:1939}. The difference between the absolute time determinations using the two models are shown in Figure~\ref{fg:absTime}. During the actual observations both models produce almost identical absolute pulse arrival times (with a discrepancy of less than 1\,ns; shown in the lower panel). The extrapolated functions converge to the timing models; however, as these models are different, the predictions of the absolute arrival times diverge. The exact method in this case for measuring the pulsar's pulse frequency and its derivative is therefore not important when interpolating timing residuals to obtain absolute pulse arrival times, but does affect extrapolated arrival times.

\section{Conclusions}

In the presence of red timing noise we recommend that our new method be used to model, interpolate or extrapolate pulsar timing noise. This method is applicable even under extreme conditions such as very steep red spectra, very large gaps and/or highly variable ToA uncertainties. No other linear unbiased interpolator or extrapolator can have lower variance than the MLE, provided that the covariance matrix is correct. We have shown that in the pulsar timing context the MLE is not unduly sensitive to errors in the covariance matrix, so its performance must be close to optimal. As data sets become longer our ability to model the timing noise will become even more important. Future applications are already being planned that will require that the timing noise be extrapolated far into the future. It will be important to estimate the error incurred by such extrapolations. The MLE estimator we have described provides a mechanism for doing so.

%Where timing data sets contain significant red noise we recommend that our new method is always used to model, interpolate or extrapolate pulsar timing noise. This method is applicable even under extreme conditions such as red noise with a very steep spectrum, large data gaps and/or highly variable ToA uncertainties. As data sets become longer be able to model the timing noise will become even more important. Future applications are already being planned that will require that the timing noise be extrapolated as far into the future as possible.

\section*{Acknowledgments} 

This work has been carried out as part of the Parkes Pulsar Timing Array project. GH is the recipient of an Australian Research Council QEII Fellowship (project \#DP0878388). The Parkes radio telescope is part of the Australia Telescope which is funded by the Commonwealth of Australia for operation as a National Facility managed by CSIRO. 

XD acknowledges support from the China Scholarship Council (CSC) which provided funding to study in Australia for one year. CSC is a non-profit institution affiliated with Ministry of Education of the P. R. China. It is entrusted by the Chinese Government with the responsibilities of managing the State Scholarship Fund and other related affairs. It sponsors Chinese citizens to pursue study abroad and international students to study in China.

\bibliography{myrefs,psrrefs,journals,modrefs}
\bibliographystyle{mn2e}

\appendix

\section{Usage instructions}

This algorithm has been implemented as part of the \textsc{interpolate} plugin for \textsc{tempo2}. Using the plugin is straightforward, but requires a model of the red timing noise. The noise is parameterised by $f_c$, $\alpha$ and $A$ according to:
\begin{equation}
P(f) = \frac{A}{\left(1+\left(\frac{f}{f_c}\right)^{\alpha/2}\right)^2}.
\end{equation}
The \textsc{spectralModel} plugin may be used to model the red noise component of the timing residuals and determine these parameters. 

Once these parameters are determined the \textsc{interpolate} plugin can be used as follows:
\begin{verbatim}
> tempo2 -gr interpolate -f mypar.par mytim.tim 
    -a A -fc fc -alpha alpha
\end{verbatim}
By default the software produces a new parameter file that includes a set of  parameters that give the interpolated function at specified times. These parameters are defined using the \textsc{IFUNC} keyword. It is also possible to provide different sampling for purposes of e.g., extrapolation using the \verb|-x1|, \verb|-x2| and \verb|-dx| command-line options.  These correspond to the start time, the last time and the step size for the extrapolation, respectively.

Figures 4 and 8 were produced using the \textsc{plotInterp} which allows the user to visualise the interpolation. Typical usage is:
\begin{verbatim}
> tempo2 -gr plotInterp -f mypar.par mytim.tim 
    -xextend 5000
\end{verbatim}
The \textsc{xextend} command line option extrapolates the prediction for 5000\,d before the first observation and 5000\,d after the last observation.
\end{document}